\documentclass{article}
\usepackage{arxiv}
\usepackage{authblk}
\pdfoutput=1

\usepackage[
backend=biber,
style=apa,
natbib,
]{biblatex}
\usepackage[hidelinks]{hyperref}

\title{AI Challenges for Society and Ethics}
\date{}
\author[1,2]{Jess Whittlestone}
\author[1]{Sam Clarke}
\affil[1]{Leverhulme Centre for the Future of Intelligence, University of Cambridge}
\affil[2]{Centre for the Study of Existential Risk, University of Cambridge}

\begin{document}
\maketitle

\begin{abstract}

Artificial intelligence is already being applied in and impacting many important sectors in society, including healthcare, finance, and policing. These applications will increase as AI capabilities continue to progress, which has the potential to be highly beneficial for society, or to cause serious harm. The role of AI governance is ultimately to take practical steps to mitigate this risk of harm while enabling the benefits of innovation in AI. This requires answering challenging empirical questions about current and potential risks and benefits of AI: assessing impacts that are often widely distributed and indirect, and making predictions about a highly uncertain future. It also requires thinking through the normative question of what beneficial use of AI in society looks like, which is equally challenging. Though different groups may agree on high-level principles that uses of AI should respect (e.g., privacy, fairness, and autonomy), challenges arise when putting these principles into practice. For example, it is straightforward to say that AI systems must protect individual privacy, but there is presumably some amount or type of privacy that most people would be willing to give up to develop life-saving medical treatments. Despite these challenges, research can and has made progress on these questions. The aim of this chapter will be to give readers an understanding of this progress, and of the challenges that remain.


\end{abstract}

\tableofcontents

\section{Introduction}

AI is already being applied in and impacting many important sectors in society, including healthcare, finance, and policing. As investment into AI research continues, we are likely to see substantial progress in AI capabilities and their potential applications, precipitating even greater societal impacts. The use of AI promises real benefits by helping us to better understand the world around us and develop new solutions to important problems, from disease to climate change. However, the power of AI systems also means that they risk causing serious harm if misused or deployed without careful consideration for their immediate and wider impacts.\footnote{ When we talk about AI systems in this chapter, we mean software systems which use machine learning (ML) techniques. ML involves learning from data to build mathematical models which can help us with a variety of real-world tasks, including predicting the likelihood a loan will be repaid based on someone’s financial history, translating text between languages, or deciding what moves to take to win at a board game.}

The role of AI governance is ultimately to take practical steps to mitigate this risk of harm while enabling the benefits of innovation in AI. To do this requires answering challenging empirical questions about the possible risks and benefits of AI, as well as challenging normative questions about what beneficial use of AI in society looks like.

To properly assess risks and benefits, we need a thorough understanding of how AI is already impacting society, and how those impacts are likely to evolve in future---which is far from straightforward. Assessing even \textit{current} impacts of a technology like AI is challenging since these are likely to be widely and variably distributed across society. Furthermore, it is difficult to determine the extent to which impacts are caused by AI systems, as opposed to other technologies or societal changes. Assessing \textit{potential} impacts of AI in the future---which is necessary if we are to intervene while impacts can still be shaped and harms have not yet occurred---is even more difficult, since it requires making predictions about an uncertain future.

The normative question of what beneficial use of AI in society looks like is also complex. A number of different groups and initiatives have attempted to articulate and agree on high-level principles that uses of AI should respect, such as privacy, fairness, and autonomy  \citep{jobin_global_2019}. Though this is a useful first step, many challenges arise when putting these principles into practice. For example, it seems straightforward to say that the use of AI systems must protect individual privacy, but there is presumably some amount or type of privacy that most people would be willing to give up to develop life-saving medical treatments. Different groups and cultures will inevitably have different views on what trade-offs we should make, and there may be no obvious answer or clear way of adjudicating between views. We must therefore also find politically feasible ways to balance different perspectives and values in practice, and ways of making decisions about AI that will be viewed as legitimate by all.

Despite these challenges, research can and has made progress on understanding the impacts of AI, and on illuminating the challenging normative questions that these impacts raise. The aim of this chapter will be to give the reader an understanding of this progress, and the challenges that remain. We begin by outlining some of the benefits and opportunities AI promises for society, before turning to some of the most concerning sources of harm and risk AI might pose. We then discuss the kinds of ethical and political challenges that arise in trying to balance these benefits and risks, before concluding with some recommendations for AI governance today.

\section{Benefits and opportunities}

The promise of AI ultimately lies in its potential to help us understand the world and solve problems more effectively than humans could do alone. We discuss potential benefits of AI in three related categories: (1) improving the quality and length of people’s lives; (2) improving our ability to tackle problems as a society; (3) enabling moral progress and cooperation.

\subsection{Improving the quality and length of people’s lives}

AI can help improve the quality and efficiency of public services and products by tailoring them to a given person or context. For example, several companies have begun to use AI to deliver personalised education resources \citep{hao_china_2019}, collecting data on students’ learning and performance and using this to better understand learning patterns and specific learning needs \citep{luan_review_2021}. Similarly, the use of AI to personalise healthcare through precision medicine---i.e. tailoring treatment based on specific features of an individual patient---is in early stages but shows real promise  \citep{xu_translating_2019,johnson_precision_2021}, with startups beginning to emerge in this space \citep{toews_these_2020}.

AI is also showing promise to drastically improve our understanding of disease and medical treatments. AI systems can now outperform human specialists on a number of specific healthcare-related tasks: for example, Google Health trained a model to predict risk of breast cancer from mammograms, which outperformed human radiologists \citep{mckinney_international_2020}. The use of AI to advance drug discovery, for instance by searching through and testing chemical compounds more quickly and effectively, is receiving increasing attention \citep{paul_artificial_2021}: the first clinical trial of an AI-designed drug began in Japan \citep{burki_new_2020} and a number of startups in this space raised substantial funds in 2020 \citep{hogarth_state_2020}. DeepMind’s AI system AlphaFold has led to substantial progress on the ``protein folding" problem,\footnote{This is the problem of predicting the 3D structure of a protein from its 2D genetic sequence.} with potential to drastically improve our ability to treat disease \citep{jumper_highly_2021}. Continued progress in AI for healthcare might even contribute to better understanding and slowing processes of ageing \citep{zhavoronkov_artificial_2019}, resulting in much longer lifespans than we enjoy today.

\subsection{Improving our ability to tackle problems as a society}

AI could help tackle many of the big challenges we face as a society, such as climate change and threats to global health, by helping model the complex systems underpinning these problems, advancing the science behind potential solutions, and improving the effectiveness of policy interventions.

For instance, AI can support early warning systems for threats such as disease outbreaks: machine learning algorithms were used to characterise and predict the transmission patterns of both Zika \citep{jiang_mapping_2018} and SARS-CoV-2 \citep{wu_nowcasting_2020,liu_deployment_2020} outbreaks, supporting more timely planning and policymaking. With better data and more sophisticated systems in future it may be possible to identify and mitigate such outbreaks much earlier \citep{schwalbe_artificial_2020}. There is also some early discussion of how AI could also be used to identify early signs of inequality and conflict: \citet{musumba_prevention_2021}, for instance, use machine learning to predict the occurrence of civil conflict in Sub-Saharan Africa. This could make it much easier to intervene early to prevent conflict.

AI-based modelling of complex systems can improve resource management, which may be particularly important in mitigating the effects of climate change. For instance, AI is beginning to see application in predicting day-ahead electricity demand in the grid, improving efficiency, and in learning how to optimally allocate resources such as fleets of vehicles to address constantly changing demand \citep{hogarth_state_2019}. Similarly, a better understanding of supply and demand in electricity grids can also help reduce reliance on high-polluting plants, and make it easier to proactively manage an increasing number of variable energy sources \citep{rolnick_tackling_2019}. Similar kinds of analysis could help with a range of other problems, including disaster response: for example, machine learning can be used to create maps from aerial imagery and retrieve information from social media to inform relief efforts  \citep{rolnick_tackling_2019}.

AI also has potential to advance science in critical areas. There are many ways that AI could improve different aspects of the scientific process: by helping us to understand and visualise patterns in data of enormous volume and dimensionality \citep{mjolsness_machine_2001,ourmazd_science_2020}; or by conducting more ‘routine’ aspects of scientific research such as literature search and summarisation, hypothesis generation, and experimental design and analysis \citep{gil_amplify_2014}. DeepMind’s work on protein folding mentioned earlier is a good example of AI already advancing science in an important area. In the future, we could see AI accelerating progress in areas like materials science, by automating the time-consuming processes in the discovery of new materials, which could help develop better materials for storing or harnessing energy, for example \citep{rolnick_tackling_2019}.

As well as improving our understanding of problems and advancing the science needed to solve them, AI can help identify the most effective solutions that currently exist. There is evidence that ML tools can be used to improve policymaking by clarifying uncertainties in data, and improving existing tools for designing and assessing interventions  \citep{rolnick_tackling_2019}. For instance, \citet{andini_targeting_2018} show that a simple ML algorithm could have been used to increase the effectiveness of a tax rebate program. It may even be possible to use AI to design more competent institutions which would help tackle many problems. One idea here is that (human) participants could determine desiderata that some institution should achieve, and leave the design of the institution to an AI system \citep{dafoe_open_2020}. This could allow novel approaches to old problems that humans cannot spot.

\subsection{Enabling moral progress and cooperation}

Most would agree that the world we live in today is a better place for most people than the world of centuries ago. This is partly due to economic and technological progress improving standards of living across the globe. But moral progress also plays an important role. Fewer people and animals experience suffering today, for example, because most people view an increasing proportion of sentient beings as worthy of care and moral concern. It has been suggested that AI could help accelerate moral progress \citep{boddington_ai_2021}, for example by playing a ``Socratic" role in helping us to reach better (moral) decisions ourselves (inspired by the role of deliberative exchange in Socratic philosophy as an aid to develop better moral judgements) \citep{lara_artificial_2020}. Specifically, such systems could help with providing empirical support for different positions, improving conceptual clarity, understanding argumentative logic, and raising awareness of personal limitations.

AI might similarly help improve cooperation between groups, which arguably underlies humans’ success in the world so far. \citet{dafoe_open_2020} outline a number of ways AI might support human cooperation: AI tools could help groups jointly learn about the world in ways that make it easier to find cooperative strategies, and more advanced machine translation could enable us to overcome practical barriers to increased international cooperation, including increased trade and possibly leading to a more borderless world. AI could also play an important role in building mechanisms to incentivise truthful information sharing, and explore the space of distributed institutions that promote desirable cooperative behaviours.

\section{Harms and risks}

Despite these many real and potential benefits, we are already beginning to see harms arise from the use of AI systems, which could become much more severe with more widespread application of increasingly capable systems.

In this section we’ll discuss five different forms of harm AI might pose for individuals and society, in each case outlining current trends and impacts of AI pointing in this direction, and what we might be especially concerned about as AI systems increase in their capabilities and ubiquity across society.

\subsection{Increasing the likelihood or severity of conflict}

AI could impact the severity of conflict by enabling the development of new and more lethal weapons. Of particular concern are lethal autonomous weapons (LAWs): systems that can select and engage targets without further intervention by a human operator, which may recently have been used in combat for the first time (UN Security Council, 2021).\footnote{According to the UN Security Council (2021) report, "Logistics convoys and retreating HAF [in Libya] were subsequently hunted down and remotely engaged by the unmanned combat aerial vehicles or the lethal autonomous weapons systems such as the STM Kargu-2 $\ldots$ programmed to attack targets without requiring data connectivity between the operator and the munition: in effect, a true ``fire, forget and find" capability."}  There is a strong case that ``armed fully autonomous drone swarms", one type of lethal autonomous weapon, qualify as a weapon of mass destruction (WMD) \citep{kallenborn_are_2020}. This means they would pose all the threats that other WMDs do: geopolitical destabilisation, and use in acts of terror or catastrophic conflict between major powers. They would also be safer to transport and harder to detect than most other WMDs \citep{aguirre_why_2020}. Beyond LAWs, AI applied to scientific research or engineering could enable the development of other extremely powerful weapons. For example, it could be used to calculate the most dangerous genome sequences in order to create especially virulent biological viruses \citep{obrien_assessing_2020,turchin_classification_2020}.

Furthermore, we are seeing more integration of AI into defense and conflict domains, which could increase the likelihood of unintentional or rapid escalation in conflict: if more military decisions are automated, this makes it harder to intervene to prevent escalation \citep{johnson_artificial_2020,deeks_machine_2018}. This is analogous to how algorithmic decision-making in financial systems led to the 2010 ‘flash crash’: automated trading algorithms, operating without sufficient oversight, caused a trillion-dollar stock market crash over a period of approximately 36 minutes. The consequences could be even worse in a conflict scenario than in finance, because there is no overarching authority to enforce failsafe mechanisms \citep{johnson_artificial_2020}.

AI could also alter incentives in a way that makes conflict more likely to occur or to escalate \citep{zwetsloot_thinking_2019}. For example, AI could undermine second strike capabilities which are central to nuclear strategic stability, by improving data collection and processing capabilities which would make it easier to discover and destroy previously secure nuclear launch facilities \citep{geist_how_2018,lieber_new_2017}.

\subsection{Making society more vulnerable to attack or accident}

As AI systems become more integral to the running of society this may create new vulnerabilities which can be exploited by bad actors. For instance, researchers managed to fool an ML model trained to recognise traffic signs into classifying a ‘stop’ sign as a ‘yield’ sign, simply by adding a small, imperceptible perturbation to the image \citep{papernot_practical_2017}. An autonomous vehicle using this model could therefore be targeted by bad actors using stickers or paint to alter traffic signs. As AI systems become more widely deployed, these kinds of attacks could have more catastrophic consequences. For example, as AI is more widely integrated into diagnostic tools in hospitals or into our transport systems, adversarial attacks could put many lives at risk \citep{finlayson_adversarial_2019,brundage_malicious_2018}.

Similarly, more widespread deployment of increasingly capable AI systems could also increase the severity of accidents. In particular, although the integration of AI into critical infrastructure has potential to bring efficiency benefits, it would also introduce the possibility of accidents on a far more consequential scale than is possible today. For example, as driverless cars become more ubiquitous, computer vision systems failing in extreme weather or road conditions could cause many cars to crash simultaneously. The direct casualties and second-order effects on road networks and supply chains could be severe. If and when AI systems become sufficiently capable to run large parts of society, these kinds of failures could possibly result in the malfunction of several critical systems at once, which at the extreme could put our very civilisation at risk of collapse.

One might think that these accidents could be avoided by making sure that a human either approves or makes the final decision. However, progress in AI capabilities such as deep reinforcement learning (DRL) could lead us to develop more autonomous systems, and there will likely be commercial pressure to deploy them. For such systems, especially when their decisions are too fast-moving or incomprehensible to humans, it is not clear how human oversight would work \citep{whittlestone_societal_2021}. 

These risks may be exacerbated by competitive dynamics in AI development. AI development is often framed in terms of a ‘race’ for strategic advantage and technological superiority between nations \citep{cave_ai_2018}. This framing is prominent in news sources, the tech sector and reports from governmental departments such as the U.S. Senate and Department of Defense \citep{imbrie_mainframes_2020}. The more AI development is underpinned by these competitive dynamics, there may be a greater incentive for actors developing in AI to underinvest in the safety and security of their systems in order to stay ahead.

\subsection{Increasing power concentration}

Several related trends suggest AI may change the distribution of power across society, perhaps drastically. 
Absent major institutional reform, it seems plausible that the harms and benefits of AI will be very unequally distributed across society. AI systems are already having discriminatory impacts on marginalised groups: for example, facial recognition software has been shown to perform many times worse for darker faces \citep{raji_actionable_2019}, and an AI system developed by Amazon to rank job candidates downgraded applications whose CVs included evidence they were female \citep{west_discriminating_2019}. Marginalised groups are less technologically literate on average, so are also more likely to be impacted by harms of AI such as the scaling up of misinformation and manipulative advertising \citep{lutz_digital_2019}. These groups are also less likely to be in a financial position to benefit from advances in AI such as personalised healthcare \citep{west_discriminating_2019}.

At the same time, AI development is making already wealthy and powerful actors more so. The companies who already have the greatest market share have access to the most data, computing power, and research talent, enabling them to build the most effective products and services---increasing their market share further and making it easier for them to continue amassing data, compute, and talent \citep{dafoe_ai_2018,kalluri_dont_2020,lee_ai_2018}. This creates a positive feedback loop cementing the powerful position these technology companies are already in. Similarly, wealthier countries able to invest more in AI development are likely to reap economic benefits more quickly than developing economies, potentially widening the gap between them. Especially if AI development leads to more rapid economic growth than previous technologies \citep{agrawal_artificial_2019}, this might cause more extreme concentration of power than we have ever seen before.

In addition, AI-based automation has the potential to drastically increase income inequality. Progress in AI systems will inevitably make it possible to automate an increasing range of tasks. Progress in reinforcement learning specifically could improve the dexterity and flexibility of robotic systems \citep{ibarz_how_2021}, leading to increased automation of manual labour jobs with lower wages. The automation of these jobs will force those people to retrain; even in the best case, they will face temporary disruptions to income \citep{lee_ai_2018}. However, it is not just low-wage or manual labour jobs that are at risk. Advances in language modelling could spur rapid automation of a wide range of knowledge work, including aspects of journalism, creative writing, and programming \citep{tamkin_understanding_2021}. Many of these knowledge workers will flood the highly social and dextrous job market (which is hard to automate, but already has low wages), further increasing income inequality  \citep{lee_ai_2018}. There is also reason to think that changes in the availability of jobs due to AI may happen more quickly than previous waves of automation, due to the fact that algorithms are infinitely replicable and instantly distributable (unlike, for example, steam engines and even computers), and the emergence of highly effect venture capital funding driving innovation  \citep{lee_ai_2018}. This gives us less time to prepare, for example by retraining those whose jobs are most likely to be lost, and makes it more likely that the impacts on inequality will be more extreme than anything seen previously.

Developments in AI are also likely to give companies and governments more control over individuals’ lives than ever before. The fact that current AI systems require large amounts of data to learn from creates incentives for companies to collect increasing amounts of personal data from users (though only certain applications such as medicine and advertising require highly personal data). Citizens are increasingly unable to consent to---or even be aware of---how their data is being used, while the collection of this data may increasingly be used by powerful actors to surveil, influence, and even manipulate and control populations. For example, the company ClearView AI scraped billions of images from Facebook, YouTube, Venmo and millions of other websites, using them to develop a ``search engine for faces'', which they then licensed, without public scrutiny, to over 600 law enforcement agencies \citep{hill_secretive_2020}. We are already seeing harmful uses of facial recognition, such as in their use to surveil Uighur and other minority populations in China  \citep{hogarth_state_2019}.\footnote{CloudWalk Technology, a key supplier to the Chinese government, markets its ``fire eye" facial recognition service to pick out ``Uighurs, Tibetans and other sensitive groups"} The simultaneous trends of apparently eroding privacy norms, and increased use of AI to monitor and influence populations, are seriously concerning.

Relatedly, AI has the potential to scale up the production of convincing yet false or misleading information online (e.g. via image, audio and text synthesis models like BigGAN and GPT-3), and to target that content at individuals and communities most likely to be receptive to it (e.g via automated A/B testing) \citep{seger_tackling_2020}. Whilst the negative impact of such techniques has so far been fairly contained, more advanced versions would make it easier for groups to seek and retain influence, for instance by influencing elections or enabling highly effective propaganda.  For example, further advances in language modelling could be applied to design tools that ``coach" their users to persuade other people of certain claims \citep{kokotajlo_persuasion_2020}. Whilst these tools could be used for social good---e.g. New York Times’ chatbot that helps users to persuade people to get vaccinated against Covid-19 \citep{gagneur_opinion_2021}---they could equally be used by self-interested groups to gain or retain influence.

\subsection{Undermining society’s ability to solve problems}

The use of AI in the production and dissemination of information online may also have broader negative impacts. In particular, it has been suggested that the use of AI to improve content recommendation engines by social media companies is contributing to worsened polarisation online \citep{ribeiro_auditing_2019,faddoul_longitudinal_2020}.\footnote{ Note that this suggestion has been disputed (e.g. \citealt{ledwich_algorithmic_2019,boxell_greater_2017}). The underlying methodological problem is that social media companies have sole access to the data required to perform a thorough analysis, and lack incentive to publicise this data or perform the analysis themselves.}

Looking to the future, the use of AI in production or targeting of information could have substantial impacts on our information ecosystem. If advanced persuasion tools are used by many different groups to advance many different ideas, we could see the world splintering into isolated ‘epistemic communities’, with little room for dialogue or transfer between them. A similar scenario could emerge via the increasing personalisation of people’s online experiences: we may see a continuation of the trend towards ``filter bubbles" and ``echo chambers", driven by content selection algorithms, that some argue is already happening \citep{barbera_tweeting_2015,flaxman_filter_2016,nguyen_exploring_2014}. In addition, increased awareness of these trends in information production and distribution could make it harder for anyone to evaluate the trustworthiness of any information source, reducing overall trust in information.

In all of these scenarios, it would be much harder for humanity to make good decisions on important issues, particularly due decreasing trust in credible multipartisan sources, which could hamper attempts at cooperation and collective action. The vaccine and mask hesitancy which exacerbated the negative impacts of Covid-19, for example, were likely the result of insufficient trust in public health advice \citep{seger_greatest_2021}. We could imagine an even more virulent pandemic, where actors exploit the opportunity to spread misinformation and disinformation to further their own ends. This could lead to dangerous practices, a significantly increased burden on health services, and much more catastrophic outcomes  \citep{seger_tackling_2020}.

\subsection{Losing control of the future to AI systems}

If AI systems continue to become more capable and begin running significant parts of the economy, we might also worry about humans losing control of important decisions. Currently, humans’ attempts to shape the world are the only goal-directed process influencing the future. However, more automated decision-making would change this, and could result in some (or all) human control over the future being lost \citep{christiano_what_2019,critch_what_2021,ngo_agi_2020,russell_human_2019}.

This concern relies on two assumptions. First, that AI systems will become capable enough that it will be not only possible but desirable to automate a majority of tasks making up the economy, from managing critical infrastructure to running corporations. Second, that despite our best efforts, we may not understand these systems well enough to be sure they are fully aligned with what their operators want.

How plausible are these assumptions? Considering the first, there is increasing reason to believe we might build AI systems as capable as humans across a broad range of economically useful tasks this century. Enormous amounts of resources are going into AI progress, and developing human-level AI is the stated goal of two very well-resourced organisations (DeepMind and OpenAI), as well as a decent proportion of AI researchers. In recent years, we have seen advances in AI defy expectations, especially in terms of their ability to solve tasks they weren’t explicitly trained for, and the improvements in performance that can be derived from simply increasing the size of models, the datasets they are trained on, and the computational resources used for training them \citep{branwen_scaling_2021}.\footnote{A similar point is made by \citet{sutton_bitter_2019}: using general methods like search and learning (rather than specific methods than involve building human knowledge into AI systems) and applying a lot of computation to them, has and will continue to yield the biggest breakthroughs in AI.} For example, GPT-3 (the latest language model from OpenAI at the time of writing), shows remarkable performance on a range of tasks it was not explicitly trained on, such as generating working code from natural language descriptions, functioning as a chatbot in limited contexts, and being used as a creative prompt \citep{tamkin_understanding_2021}. These capabilities are quickly spurring a range of commercial applications, including GitHub Copilot, a tool that helps programmers work faster by suggesting lines of code or entire functions \citep{chen_evaluating_2021}. This progress was achieved simply by scaling up previous language models to larger sizes and training them with more data and computational resources. There is good evidence that this trend will continue to result in more powerful systems without needing ‘fundamental’ breakthroughs in machine learning \citep{kaplan_scaling_2020}.

The second assumption, that advanced AI systems might not be fully aligned with or understandable to humans, is perhaps on even stronger ground. We currently train AI systems by ``trial and error", in the sense that we search for a model that does well on some objective, without necessarily knowing how a given model produces the behaviour it does. This leaves us with limited assurance about how the system might behave in new contexts or environments. A particular concern is that AI systems might help us to optimise for what we can \textit{measure} in society, but not what we actually value \citep{christiano_what_2019}. For example, we might deploy AI systems in law enforcement to help increase security and safety in communities, but later find that these systems are in fact increasing \textit{reported }sense of safety by driving down complaints and hiding information about failures. If we don’t notice these kinds of failures until AI systems are integral to the running of society, it may be very costly or even impossible to correct them. As mentioned earlier, competitive pressures to use AI for economic gain may make this more likely, driving actors to deploy AI systems without sufficient assurances that they are optimising for what we want.

This could happen gradually or suddenly, depending on the pace and shape of AI progress. The most high-profile versions of these concerns have focused on the possibility of a single misaligned AI system rapidly increasing in intelligence \citep{bostrom_superintelligence_2014}, but a much more gradual ‘takeover’ of society by AI systems may be more plausible, where humans don’t quite realise they are losing control until society is almost entirely dependent on AI systems and it is difficult or impossible for humans to regain control over decision-making.

\section{Ethical and political challenges}

It is fairly uncontroversial to suggest that improving the length and quality of people’s lives is something we should strive for, and that catastrophic accidents which claim thousands of lives should be avoided.

However, enabling the benefits of AI while mitigating the harms is not necessarily so straightforward. Sometimes what is needed to enable some area of benefit may also be the exact thing that carries risk.

For example, using AI to automate increasing amounts of the economy has potential to improve the quality of services and rapidly boost economic growth, which could result in drastic improvements to quality of life across the globe. However, the economic gains of this kind of progress, as well as the harms of job displacement, may be drastically unequal, leading to a concentration of power and rise in inequality across society never seen before. There are empirical questions here, about what processes are most likely to exacerbate inequality, that research could make progress on. There are also practical interventions that could be implemented in order to increase the likelihood that the economic gains of AI can be redistributed. However, there are still fundamental value judgements that must be made when envisioning what we want from the future of AI: how should we balance the potential for societal progress, and the possibility of huge gains in average quality of life, against the risk of radically increased inequality? If applying AI to science has potential to increase human health and lifespans, but also risks the creation of dangerous new technologies if not approached with care and wisdom, how much risk should we be willing to take? If outsourcing decisions to AI systems has potential to help us solve previously intractable societal problems, but at the cost of reduced human autonomy and understanding of the world, what should we choose?

Because these questions are normatively complex, there will be plenty of room for reasonable disagreement. Those who prioritise aggregate wellbeing will want to make different choices today to those who prioritise equality. Younger people may be happier to sacrifice privacy than older generations; those from countries which already have a strong welfare state will likely be more concerned about threats to equality; and values such as human autonomy may be perceived very differently in different cultures.

How do we deal with these disagreements? In part, this is the domain of AI ethics research, which can help to illuminate important considerations and clearly outline arguments for different perspectives. However, we should not necessarily expect ethics research to provide all the answers, especially on the timeframe in which we need to make decisions about how AI is developed and used. We can also provide opportunities for debate and resolution, but in most cases it will be impossible to resolve disagreements entirely and use AI in ways everyone agrees with.\footnote{A number of formal results from social choice theory demonstrate that when there are numerous different preferences and criteria relevant to a decision, only under strong assumptions can an unambiguously ``best" option be found - i.e. in many real-life cases, no such resolution will be possible \citep{patty_social_2014}.} We must therefore find ways to make choices about AI despite the existence of complex normative issues and disagreement on them.

Some political scientists and philosophers have suggested that where agreement on final decisions is impossible, we should instead focus our attention on ensuring the \textit{process} by which a decision is made is legitimate \citep{patty_social_2014}. This focus on decision-making procedures as opposed to outcomes has also arisen in debates around public health ethics; \citet{daniels_accountability_2008} suggest that in order to be seen as legitimate, decision-making processes must be, among other things, open to public scrutiny, revision and appeal.\footnote{Of course, there will also be room for reasonable disagreement about decision-making procedures, but we think there is likely to be less disagreement on this level, than on the level of object level decisions}

We do not currently have legitimate procedures for making decisions about how we develop and use AI in society. Many important decisions are being made in technology companies whose decisions are not open to public or even government scrutiny, meaning they have little accountability for the impacts of their decisions on society. For instance, despite being among the world’s most influential algorithms, Facebook’s and YouTube’s content selection algorithms are mostly opaque to those most impacted by them. The values and perspectives of individuals making important decisions have disproportionate influence over how ``beneficial AI" is conceived of, while the perspectives of minority groups and less powerful nations have little influence.

\section{Implications for governance}

What should we be doing to try and ensure that AI is developed and used in beneficial ways, today and in the future? We suggest that AI governance today should have three broad aims.

Ultimately, AI governance should be focused on \textbf{identifying and implementing mechanisms which enable benefits and mitigate harms of AI}. However, as we’ve discussed throughout this chapter, in some cases doing this may not be straightforward, for two reasons. First, there are many actual and potential impacts of AI which we do not yet understand well enough to identify likely harms and benefits. Second, even where impacts are well understood, tensions may arise, raising challenging ethical questions on which people with different values may disagree. AI governance therefore also needs to develop methods and processes to address these barriers: \textbf{to improve our ability to assess and anticipate the impacts of AI}; and \textbf{to make decisions even in the face of normative uncertainty and disagreement}. We conclude this chapter by making some concrete recommendations for AI governance work in each of these three categories.

\subsection{Enabling benefits and mitigating harms}

In some cases, \textbf{we might need to consider outright bans on specific applications of AI}, if the application is likely to cause a level or type of harm deemed unacceptable. For example, there has been substantial momentum behind campaigns to ban lethal autonomous weapons (LAWs),\textsuperscript{3} and the European Commission’s proposal for the first AI regulation includes a prohibition on the use of AI systems which engage in certain forms of manipulation, exploitation, indiscriminate surveillance, and social scoring \citep{noauthor_proposal_2021}. Another area where prohibitions may be appropriate is in the integration of AI systems into nuclear command and control, which could increase the risk of accidental launch with catastrophic consequences, without proportional benefits \citep{ord_future_2021}.

However, effective bans on capabilities or applications can be challenging to enforce in practice. It can be difficult to achieve the widespread international agreement needed---for example, the US government have cited the fact that China is unlikely to prohibit LAWs as justification for not making the ban themselves \citep{nscai_2021_2021}. In other cases it may be difficult to delineate harmful applications clearly enough. In the case of the EU regulation, it is likely to be very difficult to clearly determine whether an AI system should be deemed as ``manipulative" or ``exploitative" in the ways stated, for example.

Where outright bans are infeasible,\textbf{ it may be possible to limit access to powerful capabilities to reduce risk of misuse}. For example, companies might choose not to publish the full code behind specific capabilities to prevent malicious actors from being able to reproduce them, or limit access to commercial products with potential for misuse \citep{radford_better_2019}. However, this introduces a tension between the need for caution and the benefits of open sharing in promoting beneficial innovation \citep{whittlestone_tension_2020}, which has prompted substantial debate and analysis around the role of publication norms in AI \citep{gupta_report_2020}. Governments might also consider monitoring and regulating access to large amounts of computing power, which would allow them oversight and control over which actors have access to more powerful AI systems \citep{brundage_malicious_2018}.

To go beyond preventing harms and realise the full benefits of AI, it will be crucial to \textbf{invest in both socially beneficial applications, and in AI safety and responsible AI research}. Many of the potential benefits we discussed early in this chapter seem relatively underexplored: the potential uses of AI to enhance cooperation between groups, to combat climate change, or improve moral reasoning, for example, could receive a great deal more attention. Part of the barrier to working on these topics is that they may not be well-incentivised by either academia (which often rewards theoretical progress over applications) or industry (where economic incentives are not always aligned with broad societal benefit). Similarly, AI safety and responsible AI research will be crucial for ensuring even the most beneficial applications of AI do not come with unintended harms. One concrete idea would be for governments to create a fund of computational resources which is available free of charge for projects in these areas \citep{brundage_toward_2020}.

\subsection{Improving our ability to assess and anticipate impacts}

In many cases we may first need to better understand the potential impacts of AI systems before determining what kinds of governance are needed. Better and more standardised processes for impact assessment would be valuable on multiple levels.

First, we need to \textbf{establish clearer standards and methods for assuring AI systems} (also sometimes called test, evaluation, validation and verification---TEVV---methods) before they go to market, particularly in safety-critical contexts. There are currently no proven effective methods for assuring the behaviour of most AI systems, so much more work is needed \citep{flournoy_building_2020}. It is likely that rather than a single approach to assuring AI systems, an ecosystem of approaches will be needed, depending on the type of AI system, and the decision to be made \citep{ahamat_types_2021}. Better assurance processes would make it easier to decide where the use of AI systems should be restricted, by requiring uses of AI to pass certain established standards. It would also make it possible to identify and mitigate potential harms from unintended behaviour in advance, and to incentivise technical progress to make systems more robust and predictable.

\textbf{Continual monitoring and stress-testing of systems }will also be important, given it may not be possible to anticipate all possible failure modes or sources of attack in advance of deployment. Here it may be useful to build on approaches to ‘read-teaming’ in other fields including information and cyber security  \citep{brundage_malicious_2018}.

We also need \textbf{broader ways to assess and anticipate the structural impacts of AI systems}. Assurance and stress-testing can help to identify where unintended behaviours or attacks on AI systems might cause harm, but cannot identify where a system behaving as intended might nonetheless cause broader structural harms (for example, polarising online discourse or changing incentives to make conflict more likely)  \citep{zwetsloot_thinking_2019}. This will likely require looking beyond existing impact assessment frameworks and drawing on broader perspectives and methodologies, including: social science and history, fields which study how large societal impacts may come about without anyone intending them  \citep{zwetsloot_thinking_2019}; foresight processes for considering the future evolution of impacts \citep{noauthor_government_2017}; and participatory processes to enable a wider range of people to communicate harms and concerns  \citep{smith_our_2019}.

\textbf{More systematic monitoring of AI progress} would improve our ability to anticipate and prepare for new challenges before they arise \citep{whittlestone_societal_2021}. As technologies advance and more AI systems are introduced into the market, they will raise increasingly high-stakes policy challenges, making it increasingly important that governments have the capacity to react quickly. AI as a sector is naturally producing a wide range of data, metrics and measures that could be integrated into an ‘early warning system’ for new capabilities and applications which may have substantial impacts on society. Monitoring progress on widely studied benchmarks and assessment regimes in AI could enable AI governance communities to identify areas where new or more advanced applications of AI may be forthcoming. Monitoring \emph{inputs} into AI progress, such as computational costs, data, and funding, may also help to give a fuller picture of where societally-relevant progress is most likely to emerge \citep{martinez-plumed_accounting_2018}. For example, early warning signs of recent progress in language models could have been identified via a combination of monitoring progress on key benchmarks in language modelling, and monitoring the large jumps in computational resources being used to train these models.

\subsection{Making decisions under uncertainty and disagreement}

Even with better methods for assessing and anticipating the impacts of AI systems, challenges will remain: there will be uncertainties about the future impacts of AI that cannot be reduced, and conflicting perspectives on how we \textit{should }be using AI for global benefit that cannot be easily resolved. AI governance will therefore need to grapple with what \textit{processes} for making decisions about AI should look like, given this uncertainty and disagreement.

\textbf{Greater use of participatory processes} in decision-making around AI governance could help with ensuring the legitimacy and public acceptability of decisions, and may also improve the quality of the decisions themselves. There is evidence that participatory approaches used in the domain of climate policy lead to both increased engagement and understanding of decisions, and to better decisions \citep{hugel_public_2020}. Various projects have begun to engage a wider variety of perspectives in thinking through governance and societal issues related to AI \citep{noauthor_public_2017, balaram_artificial_2018}, but much more could be done, especially in terms of integrating these processes into policymaking. We would also like to see participatory studies focused on concerns and hopes about the \textit{future} of AI rather than just current AI systems, since these are more likely to be timely and relevant enough to influence decision-making. Public engagement is of course only one kind of input into decision-making processes, and must be combined with relevant expert analysis. However, participatory processes can be especially useful for understanding the wider impacts of policies which might be neglected by decision-makers, and for highlighting additional considerations or priorities, and policymaking around AI would benefit from giving them greater attention.

More generally, we need to think about how \textbf{processes for making important decisions about AI can be sufficiently open to scrutiny and challenge}. This is particularly difficult given that some of the most important decisions about the future of AI are being made within technology companies, which are not subject to the same forms of accountability or transparency requirements as governments. Some greater scrutiny may be achieved through regulation requiring greater transparency from companies. It may also be possible to improve transparency and accountability through shifts in norms---if there is enough public pressure, companies may have an incentive to be more transparent---or by improving the capacity of government to monitor company behaviour, such as by increasing technical expertise in government and establishing stronger measurement and monitoring infrastructure. 

\section{Conclusion}

In this chapter, we have outlined some of the possible ways AI could impact society into the future, both beneficial and harmful. Our aim has not been to predict the future, but to demonstrate that the possible impacts are wide-ranging, and that there are things we can do today to shape them. As well as intervening to enable specific benefits and mitigate harms, AI governance must develop more robust methods to assess and anticipate the impacts of AI, and better processes for making decisions about AI under uncertainty and disagreement.
\newpage

\printbibliography[title={Bibliography}]

\end{document}